# synthesis and modification of thin nasicon solid electrolytes using ion beams

[1]Giovanni Ceccio, [1]Jiri Vacik, [2]I. Mastronardo, [2]C. D'Urso, [1]E. Štěpanovská, [1]R. Mikšová

*[1] Department of Neutron and Ion Methods, Nuclear Physics Institute of CAS, Hlavni 130, 250 68 Řež, Prague, Czech Republic, ceccio@ujf.cas.cz*

*[2]CNR-ITAE, Via Santa Lucia sopra contesse, 5 – 98121, Messina (ME), Italy*

**Abstract**

Solid electrolytes (SEs) for sodium-based superionic conductors (NaSICON) were first introduced in 1976 and quickly recognized for their excellent ionic conductivity. While considerable effort has been made to develop thin electrolytes for all-solid-state batteries (ASSBs), only a few sodium-based SEs have been successfully fabricated as thin films. These thin films are particularly desirable for their reduced electrical resistance, which typically increases with the thickness of the SE. By reducing the thickness of the SEs to the nanometer scale, their ionic conductivity can be significantly enhanced.

In this study, the NASICON composite was initially prepared in the form of pellets using the mixed oxide technique with a planetary ball mill and synthesized by the solid-state method at 1300 °C. The resulting pellets were used as sputtering targets in a low-energy ion facility to prepare continuous NASICON nanofilms. To explore the effect of ion implantation on the electrical properties of NASICON, the prepared films were bombarded with Ni ions at 1.1 MeV and varying fluences, using the Tandetron accelerator at the CANAM infrastructure (NPI Řež). The electrical properties of both the synthesized and implanted films were analyzed through electrochemical impedance spectroscopy (EIS). The results, describing the impact of irradiation on NASICON's properties, are presented here.

## 1. INTRODUCTION

NASICON solid electrolytes, known for their high ionic conductivity, are crucial for advanced energy storage technologies [1, 2]. This study aims to enhance their performance further by improving lattice quality, systematically modifying stoichiometry through ion implantation, and investigating their behavior in thin film form. Ion beam-based techniques were employed for the production, modification, and characterization of the thin films. A systematic analysis was conducted on samples synthesized using an ion beam sputtering system and modified via ion beam implantation. Elemental composition was analyzed using ion beam-based methods, while electrochemical testing provided insights into performance improvements. The NASICON thin films were also examined to explore the impact of sample miniaturization on ionic conductivity and the potential for tuning their properties. This study compares conventional and nuclear-assisted synthesis and analytical approaches, highlighting their effects on material properties and device performance.

## 2. METHODOLOGY

The NASICON material, $Na_3Zr_2Si_2PO_{12}$ (NZSP), was produced using the solid state reaction method at ITAE, Messina. Pellets were produced using different procedures to achieve varying grain sizes in the final product. The synthesis process can be divided into three main phases: In the first phase, the precursors ($SiO_2$, $ZrO_2$, $Na_2CO_3$, $NH_4H_2PO_4$) in powder form were ball-milled to obtain a uniformly mixed fine powder. In the second phase, the fine powder was pressed into pellets and annealed at 1150°C to form the NZSP phase. In the final phase, the annealed pellets were crushed, ball-milled again, and then re-pelletized before undergoing a final sintering at 1250°C to improve the crystallinity of the NZSP pellets. Three pellets were produced: NASICON 2, made with commercial NZSP powder; NASICON 4, produced by mixing the reagents as described; and NASICON 5, prepared with the same procedure as NASICON 4, but without the second sintering step. Thin films of NASICON were then produced from these pellets using the ion beam sputtering (IBS) technique, with the pellets serving as targets for the sputtering. The ion beam sputtering was performed at the LEIF facility in NPI Řež. LEIF (Low Energy Ion Facility) is a laboratory ion beam system that generates Ar ions in the energy range of 1-100 keV and operates within the CANAM infrastructure [3]. The Ar ion beam was directed into a multipurpose vacuum chamber, where it impinged on the NASICON pellet at a 45° angle, sputtering the pellet material. Silicon wafers, placed parallel to the target (but out of the beam's direct path in a distance of about 15 cm), served as substrates. In IBS process, the sputtered material was deposited onto the substrates, forming thin NASICON films.

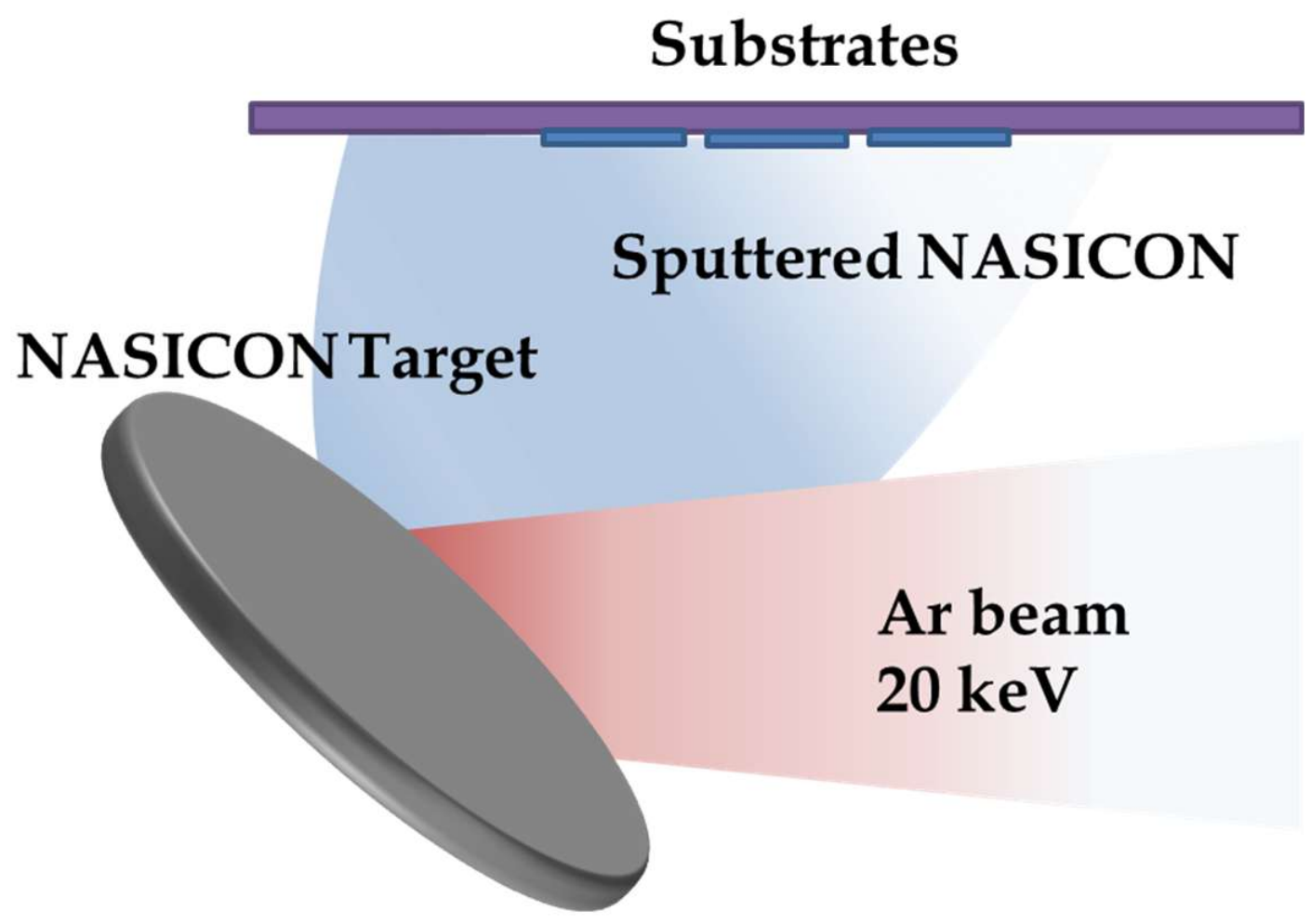

**Figure 1** Sketch of thin film deposition by ion beam sputtering

To determine the elemental composition of the prepared films, RBS measurements were performed. Given the novelty of the Ion Beam Sputtering (IBS) method for preparing NASICON ultrathin films, one of the primary goals was to determine the elemental ratios and identify any potential contaminants. RBS is widely used for studying thin films and multilayer systems, with thicknesses ranging from nanometers to micrometers. It enables the determination, with some limitations, of both the atomic mass and concentration of elemental constituents as a function of depth beneath the surface [4].

The RBS measurements were carried out at the Tandetron MC 4130 accelerator within the NPI CANAM infrastructure. In the experiment, $He^+$ ions with an energy of 2 MeV and a low current (~100 nA) were used to minimize radiation damage to the material. The backscattered particles were detected using a partially-depleted ORTEC ULTRA detector with an active area of 50 mm² and a 300 µm thick depletion layer. The detector was positioned at a backscatter angle of 170°, in accordance with the Cornell geometry.The thin films were modified by ion implantation to alter their electrical properties (ion implantation is a technique used to modify the material at shallow depths by doping it with selected atoms). Implantation was performed using 1.1 MeV Ni ions, with three different fluences (1E14 Ni/cm², 5E14 Ni/cm², and 1E15 Ni/cm²) for three different samples.

Both as-deposited and implanted samples were then investigated for changes in their electrical properties using Electrochemical Impedance Spectroscopy (EIS). The EIS study was conducted using a Biologic Potentiostat, operating in the frequency range of 7 MHz – 1 Hz. For the measurements, the samples were placed in a custom-made holder connected to the potentiostat electrodes.

## 3. RESULTS

The RBS analysis was performed on samples produced from three different pellets. The experimental data were evaluated using the SIMNRA code, which allowed for the estimation of the relative percentages of the composing elements [5]. The evaluation revealed no contaminants; however, a higher amount of oxygen was detected in pellet 5, suggesting that the stoichiometry of the NASICON layer was disturbed due to additional oxidation of the constituent elements.

**Table 1** Elemental composition of produced thin films from the RBS analysis.

| | **NASICON 2** | **NASICON 4** | **NASICON 5** |
|---|---|---|---|
| **Na (%)** | 16,39 | 15,86 | 8,83 |
| **Si (%)** | 11,11 | 10,65 | 6,67 |
| **P (%)** | 5,46 | 4,84 | 2,94 |
| **Zr (%)** | 11,29 | 10,64 | 6,62 |
| **O (%)** | 55,74 | 58,02 | 74,93 |

EIS results for the as-deposited films show that the conductivity of the films made from the commercial powder is lower than the other two, with the lowest impedance found in NASICON 4. This film, which was prepared from a pellet, subjected to two annealing steps, suggests that the observed behavior may be due to a memory effect of the original crystal structure. The impact of ion implantation on impedance is illustrated in the other images, which document the significant influence of Ni implantation. Clearly, implanted Ni initially and effectively reduces the impedance of the film, but with increasing fluence, the impedance can rise again. This behavior is due perhaps to possible competitive effects: on one hand, the increasing conductivity results from the incorporation (doping) of metal elements, while on the other hand, continued implantation leads to ion-induced damage to the crystalline structure, thus reducing the electrical conductivity.

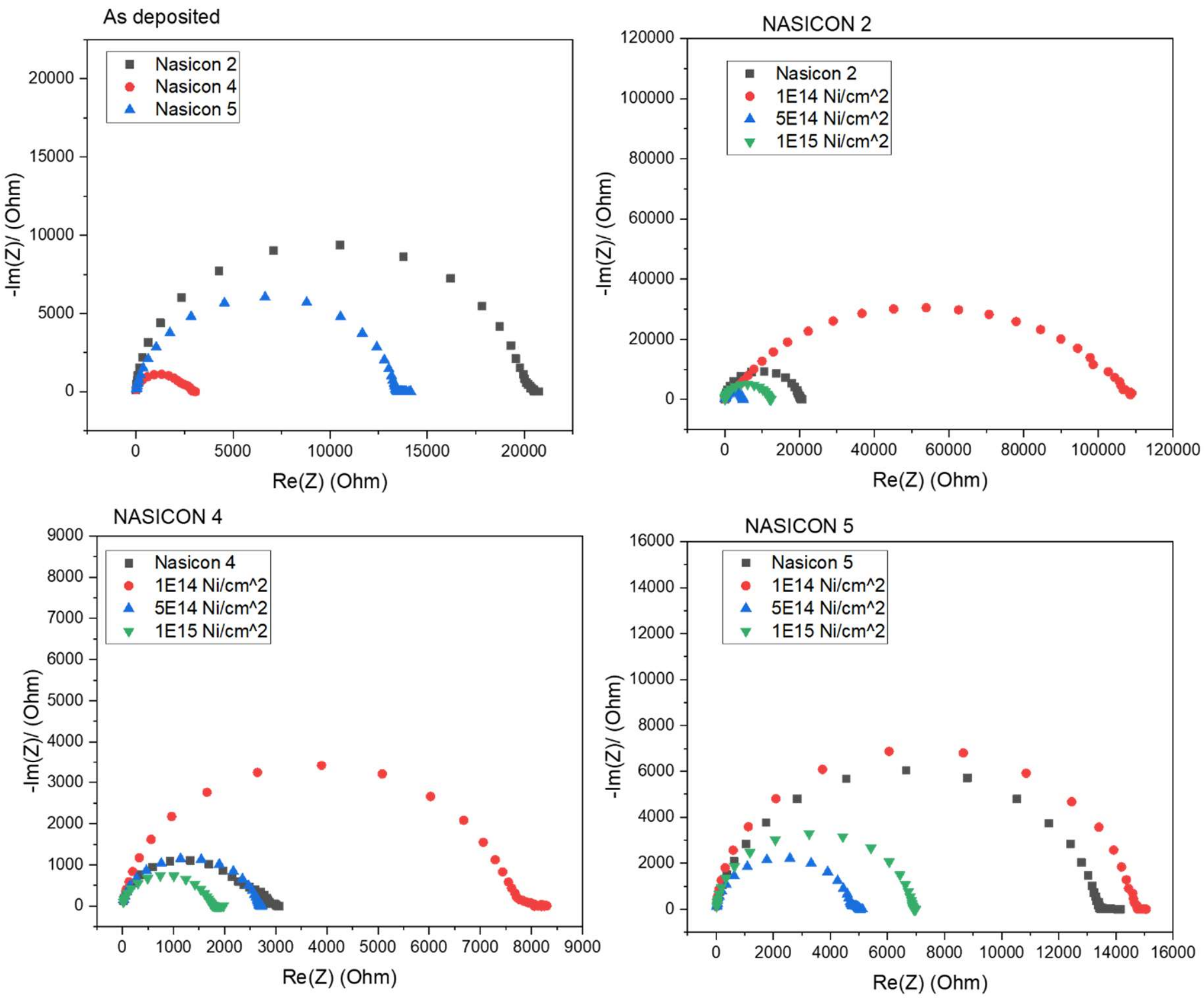

**Figure 2** Experimental impedance spectra for the NASICON thin films: a) shows the comparison between as prepared films, b), c) and d) show the changes in impedance in different NASICON after ion irradiation.

## 4. CONCLUSION

NASICON thin films were produced by ion beam sputtering from a NASICON pellet synthesized using the solid state reaction method. The films exhibited good homogeneity and no contamination (except for higher oxidation observed in one sample). The films were then irradiated with MeV Ni ions to modify their electrical properties. Electrochemical Impedance Spectroscopy (EIS) revealed that the changes in impedance were not linear with ion fluence; higher fluences also led to a reduction in the improvement of conductivity. Overall, Ni incorporation enhanced the synthesized film conductivity, significantly reducing the impedance for all NASICON films tested.

## ACKNOWLEDGEMENTS

***This work was supported by the Ministry of Education, Youth and Sports (MEYS) CR under the project OP JAK CZ.02.01.01/00/22_008/0004591. The authors acknowledged also the support of Czech Academy of Science Mobility Plus Project, Grant No. CNR-25-01.***